# Quantum dot optomechanics in suspended nanophononic strings

Anja Vogele[1], Maximilian M. Sonner[1,3], Benjamin Mayer[1], Xueyong Yuan[1,2], Matthias Weiß[1,3], Emeline D. S. Nysten1,3, Saimon F. Covre da Silva[2], Armando Rastelli[2], Hubert J. Krenner[1,3,4]*

1 Lehrstuhl für Experimentalphysik 1 and Augsburg Centre for Innovative Technologies (ACIT), Universität Augsburg, Universitätsstraße 1, 86159 Augsburg, Germany

2 Institute of Semiconductor and Solid State Physics, Johannes Kepler Universität Linz, Linz Institute of Technology, Altenbergerstraße 69, 4040 Linz, Austria

3 Nanosystems Initiative Munich (NIM), Schellingstraße 4, 80799 München, Germany

4 Center for Nanoscience (CeNS), Ludwig-Maximilians-Universität München, Geschwister-Scholl-Platz 1, 80539 München, Germany

*hubert.krenner@physik.uni-augsburg.de

The optomechanical coupling of quantum dots and flexural mechanical modes is studied in suspended nanophononic strings. The investigated devices are designed and monolithically fabricated on an (Al)GaAs heterostructure. Radio frequency elastic waves with frequencies ranging between $f = 250\,\text{MHz}$ to $400\,\text{MHz}$ are generated as Rayleigh surface acoustic waves on the unpatterned substrate and injected as Lamb waves in the nanophononic string. Quantum dots inside the nanophononic string exhibit a 15-fold enhanced optomechanical modulation compared to those dynamically strained by the Rayleigh surface acoustic wave. Detailed finite element simulations of the phononic mode spectrum of the nanophononic string confirm, that the observed modulation arises from valence band deformation potential coupling via shear strain. The corresponding optomechanical coupling parameter is quantified to $0.15\,\text{meV} \cdot \text{nm}^{-1}$. This value exceeds that reported for vibrating nanorods by approximately one order of magnitude at 100 times higher frequencies. Using this value, a derive vertical displacements in the range of $10\,\text{nm}$ is deduced from the experimentally observed modulation. The results represent an important step towards the creation of large scale optomechanical circuits interfacing single optically active quantum dots with optical and mechanical waves.

Phonons are of paramount importance for the realization of hybrid quantum architectures.[1] These fundamental excitations of condensed matter couple to almost any quantum system and experience only minute dissipation in crystalline solids. Surface acoustic waves (SAWs) are one of the very few phononic technologies of industrial relevance [2] and have recently attracted widespread interest in quantum technologies. This interest has been sparked by theoretical [3–5] work and hallmark experiments on superconducting qubits.[6–8] Semiconductor quantum dots (QDs) enable the direct transduction of the SAW phonons' radio frequencies to the optical frequencies of QD excitonic two-level system [9,10] via deformation potential and Stark effect couplings.[11–13] One of the first applications of SAW envisioned and implemented in QD-based quantum technologies was the dynamic acoustic pumping and charge state control via the acousto-electric effect.[14–17] Moreover, QDs can be integrated in fully suspended photonic crystal membranes with SAW-tunable circuit elements [18,19] enabling the dynamic control of light-matter interactions at gigahertz frequencies.[20] Such suspended systems confine both photons and phonons in the plane.[21] Interestingly, flexural, anti-symmetric Lamb modes excited in these membranes are stress-neutral in the center plane of the membrane, i.e. the volumetric strain vanishes. Thus, so far experiments focusing on the optomechanical coupling between QDs and mechanical excitations used samples in which the dots were deliberately displaced from the membrane's center.[22–24] In contrast, further experiments showed an unexpected pronounced tuning of QDs placed in the center of the membrane [20] indicating strong optomechanical coupling of the QD exciton with modulation amplitudes comparable to that observed of a high-Q cavity mode.

In this communication, we report on the optomechanical coupling of single GaAs/(Al)GaAs QDs to the phononic modes of suspended nanophononic strings. In the studied frequency band from $f = 250 \text{ MHz}$ to $400 \text{ MHz}$, we observe large tuning amplitudes which are enhanced by more than a factor of 15 compared to those of QDs strained by conventional Rayleigh SAWs propagating on the unpatterned surface.

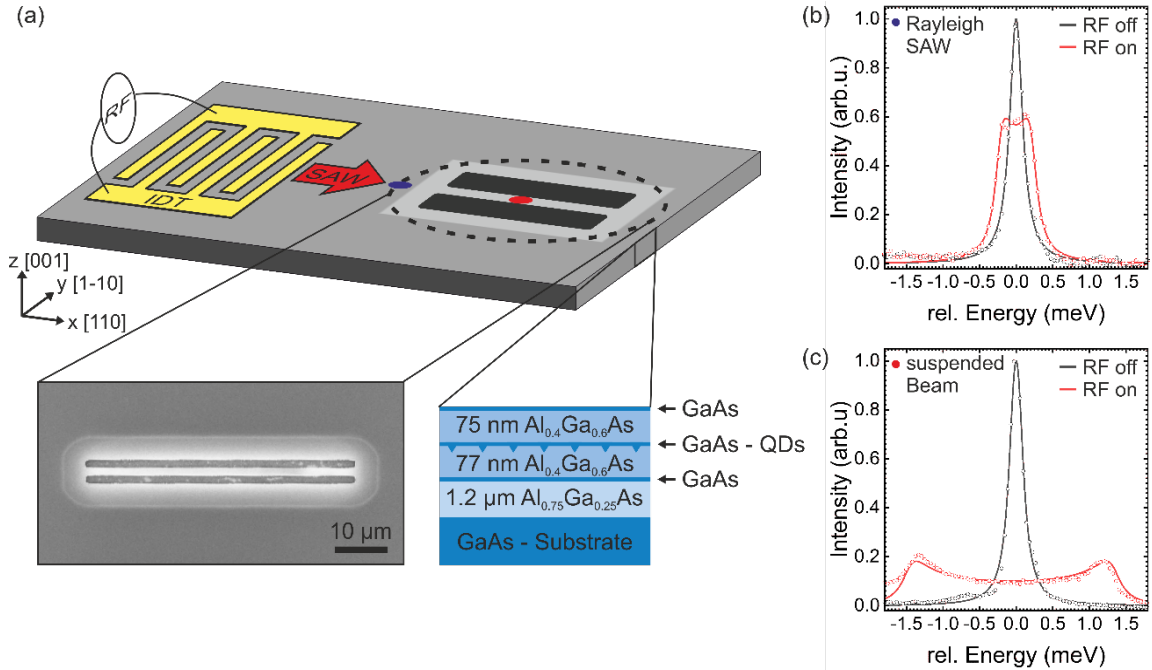

**Figure 1.** (a) Sample layout consisting of a metal IDT and a 50 μm-long and 2 μm-wide suspended nanophononic string fabricated on an (Al,Ga)As-based heterostructure. A scanning electron microscope image of a typical nanophononic string is shown below. (b), (c) Emission of a single QD in the unpatterned region and inside of a nanophononic string (marked by the blue and red dot in the layout) without/with RF voltage applied to the IDT (black/red) at $f_{RF} = 370$ MHz. Solid lines are best fits of a time modulated Lorentzian (Equation 1) to the experimental data.

Devices were fabricated on a (Al)GaAs heterostructure containing a single layer of GaAs QDs embedded in (Al)GaAs barriers in its center. This type of QD is particularly suited to study strain tuning by elastic waves[25] since charging due to the acoustoelectric effect [16,26] is almost completely suppressed. **Figure 1** (a) shows a schematic of the full device comprising a chirped interdigital transducer, labelled IDT, and a 50 μm-long suspended nanophononic string. A scanning electron microscope image of a typical nanophononic string and the full layer sequence of the heterostructure are shown as well. Details on device fabrication can be found in the experimental section. When applying a radio frequency (RF) voltage to the IDT a SAW is generated. Figure 1 (b) and (c) compare the optomechanical response of two QDs in the unpatterned region [marked by blue dot in (a)] and inside of a nanophononic string [marked by red dot in (a)], respectively. We plot the measured photon energy, $E$, relative to the center energy of the unmodulated line, $E_0$, as symbols. The data plotted in red (black) are the emission intensity of a QD transition when the RF signal is switched on (off) and the QD is (not) dynamically strained. Clearly, the emission lines show the expected broadening in at both positions when the RF signal ($f_{RF} = 370 \text{ MHz}$, $P_{RF} = 29 \text{ dBm}$) is switched on. Most strikingly, the broadening observed for the QD inside the nanophononic string is largely enhanced compared to the QD in the unpatterned region. The observed broadening can be described by a Lorentzian line modulated in time with a frequency $f_{RF}$.[27–29] It is given by

$$I(E) = I_0 + f_{RF} \frac{2A}{\pi} \int_0^{1/f_{rf}} \frac{w}{4 \cdot (E-(E_0+\Delta E \cdot \sin(2\pi \cdot f_{RF} \cdot t)))^2 + w^2} dt, \quad (1)$$

in which $\Delta E$ and $w$ denote the amplitude of the optomechanical modulation and the unperturbed linewidth of the QD emission line, respectively. Best fits of **Equation 1** (full lines in Figure 1 (b) and (c)) faithfully reproduce the experimental data and allow us to quantify $\Delta E$ to be $0.23 \text{ meV}$ and $1.40 \text{ meV}$ for the Rayleigh SAW and on the nanophononic string, respectively.

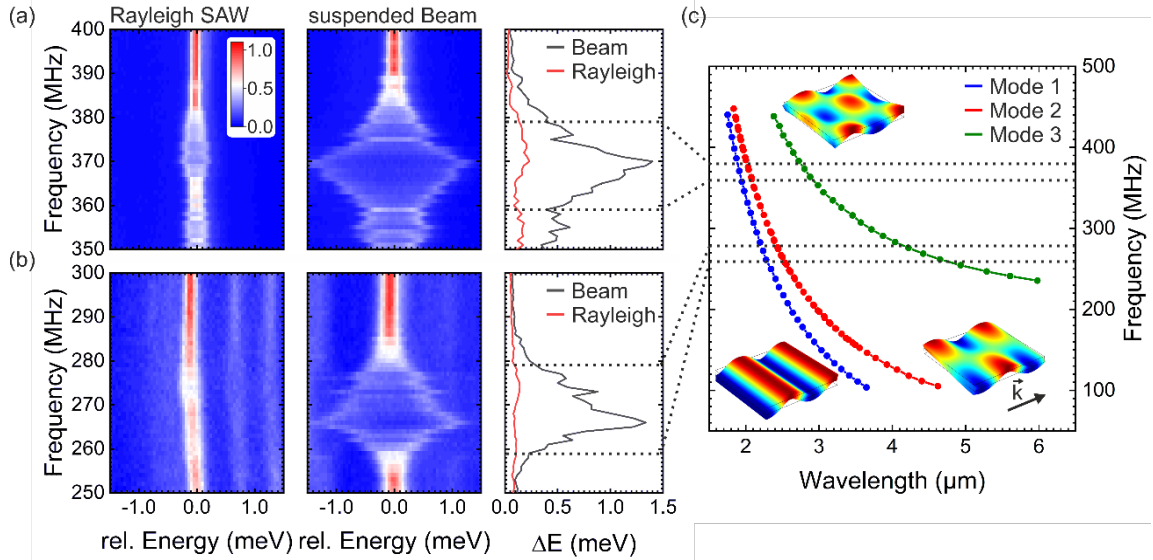

**Figure 2.** a), b) False-color plots of the normalized emission intensity of QDs modulated by a Rayleigh-type SAW (left panel) and inside a nanophononic string (center panel) as a function of $f_{RF}$ and $\Delta E$. Right Panel: Optomechanical modulation amplitude $\Delta E$ derived from the data in the other panels. The QDs embedded in a nanophononic string show an enhanced optomechanical coupling compared to the QD in the unpatterned region. c) Simulated dispersion of flexural modes of a 50 μm-long and 2 μm-wide suspended nanophononic string in the $f_{RF}$ range covered in our experiment. Each data point represents one calculated stationary mode. The dotted lines mark the frequency ranges in which pronounced coupling is observed. The displacement profiles of the calculated modes are shown in the insets

Next, we analyze the frequency dependence of the optomechanical coupling in both areas. **Figure 2** (a) and (b) shows $f_{RF}$-dependent emission spectra of QDs strained by a Rayleigh wave (left panels) and inside a nanophononic string (right panels). The data in Figure 2 (a) and (b) are recorded with a RF signal ($P_{RF} = 29$ dBm) applied by two different IDTs with designs nominally facilitating SAW excitation from $250 - 300$ MHz and $350 - 400$ MHz, respectively. The normalized emission intensity is color-coded, with blue (red) corresponding to minimum (maximum) intensity, and plotted as a function of $f_{RF}$ (vertical axis) and relative emission energy $E - E_0$ (horizontal axis). As $f_{RF}$ is tuned, both QDs in the unpatterned region shown in Figure 2 (a) and (b) exhibit weak optomechanical modulations. The modulations do not cover the full nominal frequency ranges of the IDTs due to imperfections in the nanofabrication and the limited number of finger pairs. In contrast to QDs strained by Rayleigh waves, the two dots inside the nanophononic string show strong enhancement of the optomechanical modulation over a wide frequency band. For the two frequency bands studied we deduce enhancements by factors of 10 and 15 with respect to the corresponding data obtained with Rayleigh SAWs, as shown in Figure 2 (a) and (b), respectively. Moreover, we detect a strong modulation between $260 - 270$ MHz and $360 - 380$ MHz, where no modulation is observed for the QDs in the unpatterned region. Remarkably, for the frequency band studied in Figure 2 (b) the largest modulation is observed for the QD in the nanophononic string at $f_{RF} = 266$ MHz. The extracted $\Delta E$ are plotted as a function of $f_{RF}$ in the right panels, demonstrating the broadband enhancement of the optomechanical modulation inside the nanophononic string. We performed finite element model (FEM) simulation of the phononic modes of the string using a geometry derived from our SEM characterization. The obtained dispersion in our 50 µm-long and 2 µm-wide string is plotted in Figure 2 (c). Our simulation confirms the propagation of three flexural modes which we refer to as Mode 1, Mode 2 and Mode 3. The dispersions are plotted in red, blue and green in Figure 2 (c). Mode 1 and Mode 2 are the two fundamental flexural modes which exhibit an antinode and a node in the center of the beam along its major axis, respectively. Mode 3 is the first higher order mode. In addition to propagating solutions, a series of bound resonances, which are marked by symbols overlapped over the curves in Figure 2 (c). In our experiment we couple Rayleigh SAWs into these modes and, therefore, we expect that we excite all three modes both as bound resonances as well as and propagating modes.[30,31] The excitation of propagating flexural modes is corroborated by stroboscopic spectroscopy discussed in the supporting information. Furthermore, in the frequency band in which we observe strong enhancement of the optomechanical modulation (marked by dashed lines), our FEM simulations in Figure 2 (c) predict multiple bound resonances for all three modes. Hence, individual resonances apparent in the data shown in the center and right panels of Figure 2 (a) and (b) cannot be unambiguously attributed to certain bound resonances. Moreover, we expect a strong variation of the coupling efficiency for different frequencies due to two additional effects. First, the frequency response of the IDT is not flat over the full range of frequencies and, second, the impedance mismatch between the Rayleigh SAW and Lamb waves. [30] Nevertheless, our FEM simulations allow us to quantify the strength of optomechanical coupling for the three flexural modes. Since the center of the beam at which the layer of QDs is located is a stress-neutral plane for flexural modes, the volumetric strain vanishes. Thus, the observed optomechanical response of the QDs has to arise from off-diagonal elements of the strain tensor (shear strain) and the corresponding deformation potentials. Deformation potential coupling by shear strain is known to be weak compared to that by the diagonal elements (normal strain). Thus, the investigation of deformation potential coupling due to shear strain is typically challenging and our experiment elegantly suppresses the dominant normal strains.

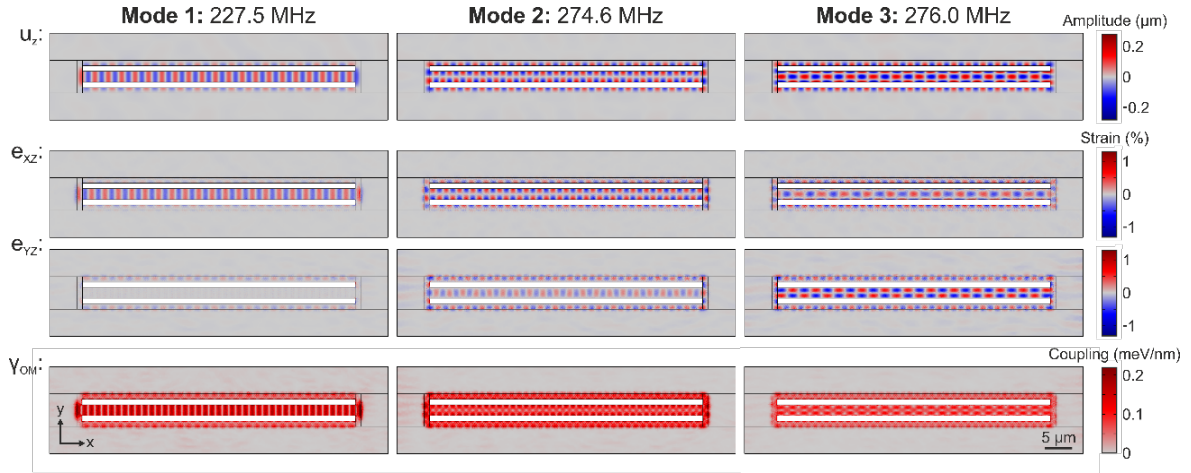

**Figure 3.** Simulated flexural mode profiles of the nanophononic string in the QD layer calculated by FEM. The vertical displacements $u_z$, shear strains components $e_{xz}$ and $e_{yz}$ as well as optomechanical coupling parameter $\gamma_{OM}$ are depicted for the three mode profiles. As the volumetric strain vanishes at the QD layer, only shear strains of the modes contribute to the observed optomechanical tuning.

In **Figure 3** we analyze bound resonances of each of the three different modes. The upper rows show the vertical displacements ($u_z$) which confirm the different symmetries of the three modes. The shear strains $e_{xz}$ and $e_{yz}$ are plotted in the two center rows. For Mode 1, only $e_{xz}$ is finite, while for Mode 2 and Mode 3 both shear strains contribute to the observed optomechanical tuning. Using the Pikus-Bir strain Hamiltonian and considering that the conduction band is negligibly affected by shear strains,[32] we determine the optomechanical coupling parameter to

$$\gamma_{om} = \frac{\partial E_{QD}}{\partial u_z} = d \cdot \frac{\sqrt{(e_{xz}^2 + e_{yz}^2)}}{u_z}. \qquad (2)$$

Because the center of the nanostring is a stress-neutral plane, **Equation 2** only contains the valence band deformation potential $d$ and the shear strain tensor components $e_{xz}$ and $e_{yz}$. Using our FEM simulations and Equation 2 we are able to directly evaluate $\gamma_{om}$ for all three modes using the bulk value of the deformation potential of GaAs, $d = -4.8$ eV. [33] The obtained profiles are plotted in the bottom row of Figure 3. Our simulations predict the largest $\gamma_{om} \approx 0.2$ meV/nm for Mode 1 and $\gamma_{om} \approx 0.1 - 0.15$ meV·nm$^{-1}$ for Mode 2 and Mode 3. As expected, these values are approximately one order of magnitude smaller than that for Rayleigh waves, for which strong normal (volumetric) strain dominates.[9,25] However, the values obtained are about one order of magnitude larger than that reported for vibrating nanowires[27,34] at more than two orders of magnitude higher frequencies. From our simulations we can also quantify the displacements corresponding to the values of $\Delta E$ observed in our experiment. The maxima $\Delta E = 1.401$ meV at $f_{RF} = 370$ MHz and $1.336$ meV at $f_{RF} = 266$ MHz in Figure 2 (a) and (b) correspond to $u_z = 10.5 \pm 3.5$ nm and $10.0 \pm 3.4$ nm, respectively. Analogous analysis of the data for Rayleigh waves yields $u_z = 0.08$ nm and $0.04$ nm at these frequencies. The strong enhancement of $u_z$ in the nanophononic string is a direct consequence of the localization of mechanical energy in this fully suspended structure and that the string can freely oscillate, in contrast to the Rayleigh wave.

In conclusion, we demonstrate that QDs can be coupled to flexural modes of a suspended nanophononic string and observe a strong enhancement of the optomechanically induced spectral modulation at radio frequencies exceeding 400 MHz, important to reach the resolved sideband regime.[10] In this regime parametric transduction becomes accessible and enables the implementation of hybrid quantum dot optomechanical transduction and control schemes. Furthermore, our scheme can be directly applied to spin qubits of optically active defect centers,[35–37] for which recent proposals[38–40] promise high fidelity quantum control schemes, or QDs forming in nanowires.[12,13,17,41,42] Moreover, our work marks a first important step to interface optomechanical crystals with engineered dispersions of phonons and photons and operation frequencies in the GHz domain.[43,44] Finally, the observed optomechanical coupling arises exclusively from shear strain modulating the valence band of the semiconductor, an rarely studied effect compared to normal strain coupling.

**Experimental Section**

*Sample design*:

Nanophononic strings were fabricated on an (Al)GaAs heterostructure consisting of the following layer sequence (beginning from the GaAs substrate): a 1.2 μm thick $Al_{0.8}Ga_{0.2}As$ sacrificial layer, a 4 nm thick GaAs layer followed by a 77 nm thick $Al_{0.4}Ga_{0.6}As$ layer, the GaAs QD layer obtained by filling with 2 nm GaAs droplet-etched nanoholes, a 75 nm thick $Al_{0.4}Ga_{0.6}As$ layer, and a 4 nm GaAs think layer. The beam pattern was defined by electron beam lithography and transferred using ICP-RIE using a $BCl_3/Cl_2/Ar$ process and undercut was obtained by hydrofluoric acid (HF) and critical point drying. After underetching, the QDs are located in the middle of the nanobeam along the growth direction and the two 4 nm thick GaAs layers protect the Al-containing strings from oxidation. Cr/Au (5 nm/ 50 nm) multi-passband IDTs were fabricated in a standard lift-off process.[29]

*Acousto-optical spectroscopy*:

QDs are studied by conventional low temperature ($T = 10 \text{ K}$) microphotoluminescence (µ-PL) spectroscopy. Rayleigh SAWs are generated on the unpatterned heterostructure by a signal generator connected to the IDTs. The SAW is generated in short pulses to suppress unwanted heating of the sample.[20] Furthermore, the laser exciting the µ-PL is synchronized with the electrically generated SAW pulses by a delay generator to probe the QDs only when the acoustic wave is present. In stroboscopic experiments the laser repetition rate was commensurate to $f_{RF}$ and the relative phase was tuned.[18,45]

*Numerical simulations*:

The phononic mode spectrum was calculated using COMSOL Multiphysics using a tetrahedral mesh and bulk mechanical properties of all materials of the heterostructure. The geometry for simulations comprises in vertical direction the nominal heterostructure and in the plane the shape of the string and the adjacent undercut area as derived from optical microscope and SEM images.

**Supporting Information**
Supporting Information is available from the Wiley Online Library or from the author.

**Acknowledgements**
We gratefully acknowledge support by Deutsche Forschungsgemeinschaft (DFG, German Research Foundation) via the German Excellence Initiative's Cluster of Excellence "Nanosystems Initiative Munich" (NIM) and KR3790/6-1, the Austrian Science Fund (FWF) P29603, the Linz Institute of Technology (LIT) and the LIT Secure and Correct Systems Lab supported by the state of Upper Austria. We thank Achim Wixforth for his enduring support and invaluable discussions.

References

[1] G. Kurizki, P. Bertet, Y. Kubo, K. Mølmer, D. Petrosyan, P. Rabl, and J. Schmiedmayer, Proc. Natl. Acad. Sci. U. S. A. **112**, 3866 (2015).

[2] P. Delsing, A.N. Cleland, M.J.A. Schuetz, J. Knörzer, G. Giedke, J.I. Cirac, K. Srinivasan, M. Wu, K.C. Balram, C. Bäuerle, T. Meunier, C.J.B. Ford, P. V Santos, E. Cerda-Méndez, H. Wang, H.J. Krenner, E.D.S. Nysten, M. Weiß, G.R. Nash, L. Thevenard, C. Gourdon, P. Rovillain, M. Marangolo, J.-Y. Duquesne, G. Fischerauer, W. Ruile, A. Reiner, B. Paschke, D. Denysenko, D. Volkmer, A. Wixforth, H. Bruus, M. Wiklund, J. Reboud, J.M. Cooper, Y. Fu, M.S. Brugger, F. Rehfeldt, and C. Westerhausen, J. Phys. D. Appl. Phys. **52**, 353001 (2019).

[3] R. Blattmann, H.J. Krenner, S. Kohler, and P. Hänggi, Phys. Rev. A **89**, 012327 (2014).

[4] M.J.A. Schuetz, E.M. Kessler, G. Giedke, L.M.K. Vandersypen, M.D. Lukin, and J.I. Cirac, Phys. Rev. X **5**, 031031 (2015).

[5] M.J.A. Schuetz, J. Knörzer, G. Giedke, L.M.K. Vandersypen, M.D. Lukin, and J.I. Cirac, Phys. Rev. X **7**, 041019 (2017).

[6] M. V. Gustafsson, T. Aref, A.F. Kockum, M.K. Ekstrom, G. Johansson, and P. Delsing, Science **346**, 207 (2014).

[7] A. Bienfait, K.J. Satzinger, Y.P. Zhong, H.-S. Chang, M.-H. Chou, C.R. Conner, É. Dumur, J. Grebel, G.A. Peairs, R.G. Povey, and A.N. Cleland, Science **364**, 368 (2019).

[8] B.A. Moores, L.R. Sletten, J.J. Viennot, and K.W. Lehnert, Phys. Rev. Lett. **120**, 227701 (2018).

[9] M. Weiß and H.J. Krenner, J. Phys. D. Appl. Phys. **51**, 373001 (2018).

[10] M. Metcalfe, S.M. Carr, A. Muller, G.S. Solomon, and J. Lawall, Phys. Rev. Lett. **105**, 37401 (2010).

[11] J.R. Gell, M.B. Ward, R.J. Young, R.M. Stevenson, P. Atkinson, D. Anderson, G.A.C. Jones, D.A. Ritchie, and A.J. Shields, Appl. Phys. Lett. **93**, 81115 (2008).

[12] M. Weiß, J.B. Kinzel, F.J.R. Schülein, M. Heigl, D. Rudolph, S. Morkötter, M. Döblinger, M. Bichler, G. Abstreiter, J.J. Finley, G. Koblmüller, A. Wixforth, and H.J. Krenner, Nano Lett. **14**, 2256 (2014).

[13] M. Weiß, F.J.R. Schülein, J.B. Kinzel, M. Heigl, D. Rudolph, M. Bichler, G. Abstreiter, J.J. Finley, A. Wixforth, G. Koblmüller, and H.J. Krenner, J. Phys. D. Appl. Phys. **47**, 394011 (2014).

[14] C. Wiele, F. Haake, C. Rocke, and A. Wixforth, Phys. Rev. A **58**, R2680 (1998).

[15] O.D.D. Couto, S. Lazić, F. Iikawa, J.A.H. Stotz, U. Jahn, R. Hey, and P. V. Santos, Nat. Photonics **3**, 645 (2009).

[16] F.J.R. Schülein, K. Müller, M. Bichler, G. Koblmüller, J.J. Finley, A. Wixforth, and H.J. Krenner, Phys. Rev. B **88**, 085307 (2013).

[17] A. Hernández-Mínguez, M. Möller, S. Breuer, C. Pfüller, C. Somaschini, S. Lazić, O. Brandt, A. García-Cristóbal, M.M. de Lima, A. Cantarero, L. Geelhaar, H. Riechert, and P. V Santos, Nano Lett. **12**, 252 (2012).

[18] D.A. Fuhrmann, S.M. Thon, H. Kim, D. Bouwmeester, P.M. Petroff, A. Wixforth, and H.J. Krenner, Nat. Photonics **5**, 605 (2011).

[19] S. Kapfinger, T. Reichert, S. Lichtmannecker, K. Müller, J.J. Finley, A. Wixforth, M. Kaniber, and H.J. Krenner, Nat. Commun. **6**, 8540 (2015).

[20] M. Weiß, S. Kapfinger, T. Reichert, J.J. Finley, A. Wixforth, M. Kaniber, and H.J. Krenner, Appl. Phys. Lett. **109**, 033105 (2016).

[21] E. Gavartin, R. Braive, I. Sagnes, O. Arcizet, A. Beveratos, T.J. Kippenberg, and I. Robert-Philip, Phys. Rev. Lett. **106**, 203902 (2011).

[22] S.G. Carter, A.S. Bracker, M.K. Yakes, M.K. Zalalutdinov, M. Kim, C.S. Kim, C. Czarnocki, M. Scheibner, and D. Gammon, Appl. Phys. Lett. **111**, 183101 (2017).

[23] S.G. Carter, A.S. Bracker, G.W. Bryant, M. Kim, C.S. Kim, M.K. Zalalutdinov, M.K. Yakes, C. Czarnocki, J. Casara, M. Scheibner, and D. Gammon, Phys. Rev. Lett. **121**, 246801 (2018).

[24] X. Yuan, M. Schwendtner, R. Trotta, Y. Huo, J. Martín-Sánchez, G. Piredda, H. Huang, J. Edlinger, C. Diskus, O.G. Schmidt, B. Jakoby, H.J. Krenner, and A. Rastelli, Arxiv:1905.07738 (2019).

[25] F.J.R. Schülein, E. Zallo, P. Atkinson, O.G. Schmidt, R. Trotta, A. Rastelli, A. Wixforth, and H.J. Krenner, Nat. Nanotechnol. **10**, 512 (2015).

[26] S. Völk, F.J.R. Schülein, F. Knall, D. Reuter, A.D. Wieck, T.A. Truong, H. Kim, P.M. Petroff, A. Wixforth, and H.J. Krenner, Nano Lett. **10**, 3399 (2010).

[27] I. Yeo, P.-L. de Assis, A. Gloppe, E. Dupont-Ferrier, P. Verlot, N.S. Malik, E. Dupuy, J. Claudon, J.-M. Gérard, A. Auffèves, G. Nogues, S. Seidelin, J.-P. Poizat, O. Arcizet, and M. Richard, Nat. Nanotechnol. **9**, 106 (2013).

[28] E.D.S. Nysten, Y.H. Huo, H. Yu, G.F. Song, A. Rastelli, and H.J. Krenner, J. Phys. D. Appl. Phys. **50**, 43LT01 (2017).

[29] M. Weiß, A.L. Hörner, E. Zallo, P. Atkinson, A. Rastelli, O.G. Schmidt, A. Wixforth, and H.J. Krenner, Phys. Rev. Appl. **9**, 014004 (2018).

[30] A. V Korovin, Y. Pennec, M. Stocchi, D. Mencarelli, L. Pierantoni, T. Makkonen, J. Ahopelto, and B. Djafari Rouhani, J. Phys. D. Appl. Phys. **52**, 32LT01 (2019).

[31] Y. Takagaki, E. Wiebicke, P. V Santos, R. Hey, and K.H. Ploog, Semicond. Sci. Technol. **17**, 1008 (2002).

[32] P.Y. Yu and M. Cardona, *Fundamentals of Semiconductors: Physics and Material Properties* (Springer, Berlin, 2005).

[33] I. Vurgaftman, J.R. Meyer, and L.R. Ram-Mohan, J. Appl. Phys. **89**, 5815 (2001).

[34] M. Montinaro, G. Wüst, M. Munsch, Y. Fontana, E. Russo-Averchi, M. Heiss, A. Fontcuberta i Morral, R.J. Warburton, and M. Poggio, Nano Lett. **14**, 4454 (2014).

[35] D.A. Golter, T. Oo, M. Amezcua, I. Lekavicius, K.A. Stewart, and H. Wang, Phys. Rev. X **6**, 041060 (2016).

[36] S.J. Whiteley, G. Wolfowicz, C.P. Anderson, A. Bourassa, H. Ma, M. Ye, G. Koolstra, K.J. Satzinger, M. V. Holt, F.J. Heremans, A.N. Cleland, D.I. Schuster, G. Galli, and D.D. Awschalom, Nat. Phys. **15**, 490 (2019).

[37] F. Iikawa, A. Hernández-Mínguez, I. Aharonovich, S. Nakhaie, Y.-T. Liou, J.M.J. Lopes, and P. V. Santos, Appl. Phys. Lett. **114**, 171104 (2019).

[38] G. Calajó, M.J.A. Schuetz, H. Pichler, M.D. Lukin, P. Schneeweiss, J. Volz, and P. Rabl, Phys. Rev. A **99**, 053852 (2019).

[39] M.-A. Lemonde, S. Meesala, A. Sipahigil, M.J.A. Schuetz, M.D. Lukin, M. Loncar, and P. Rabl, Phys. Rev. Lett. **120**, 213603 (2018).

[40] M.C. Kuzyk and H. Wang, Phys. Rev. X **8**, 041027 (2018).

[41] S. Lazić, E. Chernysheva, A. Hernández-Mínguez, P. V Santos, and H.P. van der Meulen, J. Phys. D. Appl. Phys. **51**, 104001 (2018).

[42] S. Lazić, E. Chernysheva, Ž. Gačević, H.P. van der Meulen, E. Calleja, and J.M. Calleja Pardo, AIP Adv. **5**, 097217 (2015).

[43] M. Eichenfield, J. Chan, R.M. Camacho, K.J. Vahala, and O. Painter, Nature **462**, 78 (2009).

[44] K.C. Balram, M.I. Davanço, J.D. Song, and K. Srinivasan, Nat. Photonics **10**, 346 (2016).

[45] S. Völk, F. Knall, F.J.R. Schülein, T.A. Truong, H. Kim, P.M. Petroff, A. Wixforth, and H.J. Krenner, Appl. Phys. Lett. **98**, 23109 (2011).

# Supporting Information

**Quantum dot optomechanics in suspended nanophononic strings**

*Anja Vogele, Maximilian M. Sonner, Benjamin Mayer, Xueyong Yuan, Matthias Weiß, Emeline D. S. Nysten, Saimon F. Covre da Silva, Armando Rastelli, Hubert J. Krenner \**

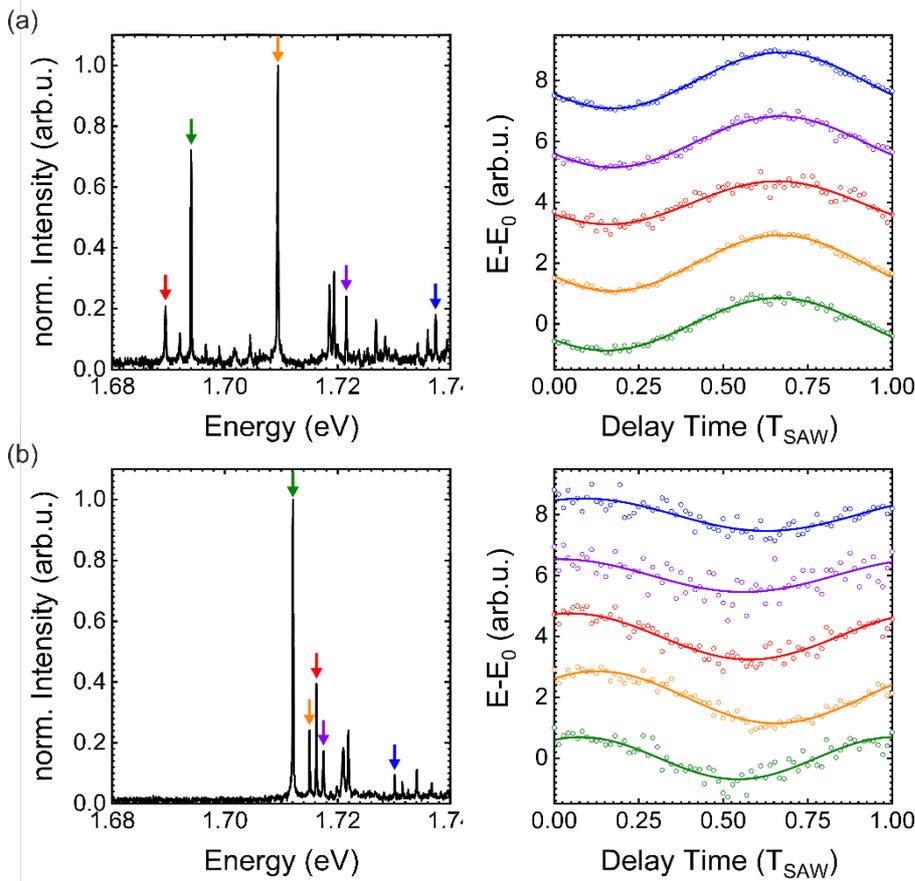

**Figure S1.** Stroboscopic photoluminescence spectroscopy of QDs in the unpatterned region (a) and on the nanophononic string (b). Left panels: Emission spectra of the QDs without RF applied. Right Panels: Normalized spectral shift of the QDs with RF applied as a function of delay times between laser and SAW excitation (symbols) and best fit of a sinusoidal modulation (lines). Colors correspond to QD emission line marked in the spectrum by an arrow of the same color.

Figure S1 shows stroboscopic photoluminescence spectroscopy[45] of QDs in the unpatterned region (a) and on the nanophononic string [laser repetition rate $f_{Laser} = 70 \text{ MHz}$, $f_{RF} = 280 \text{ MHz} = 4 \cdot f_{Laser}$] (b). Emission spectra of the QDs without RF applied are presented in the left panels with the emission lines originating from several spatially separated QDs within the region of the diffraction limited laser spot (diameter $\approx$ 1.5 μm). In the right panels, the measured normalized spectral shift of the QDs with RF applied, $E - E_0$ ($E_0$ being the center energy of the unperturbed emission peak) is plotted as symbols as a function of delay time between laser and SAW excitation. Arrows indicate the corresponding QD emission line in the left panel. These stroboscopic measurements allow for the excitation of the QDs at a fixed local phase of the SAW, i.e. at fixed time during the SAW's oscillation by tuning the delay time. All measured spectral modulations (symbols) are faithfully reproduced by best fits of a sinusoidal modulation (lines), as expected for deformation potential coupling.[12] For the Rayleigh wave, all QDs exhibit the same phase dependent modulation of $E - E_0$ as shown in the right panel of Figure S1 (a). This is expected as the wavelength of a Rayleigh wave on the unpatterned region is much larger than the diameter of the diffraction-limited laser spot, so that the local phase of the SAW is approximately the same for all QDs at every delay time during the scan. Hence, all QDs exhibit the same phase relationship.

On the nanophononic string, the phase dependences of $E - E_0$ of investigated QDs exhibit clear phase shifts as can be seen in the right panel of Figure S1 (b). This behavior is precisely expected for propagating antisymmetric Lamb waves on the nanophononic string because the wavelength is comparable to the spot size of the laser [see Figure 2 (c)]. As a result, the phase varies across the diffraction limited laser spot at a fixed delay time. Thus, the phase dependences of individual emission lines are different as observed in Figure S1(b).

This observation excludes the *exclusive* excitation of standing waves and symmetric Lamb waves. For the first, nodes are fixed and do not show any modulation. In the two neighboring regions, points oscillate in phase within each region and relative with a $\pi$ phaseshift. For the latter, the wavelength is greater than that of the Rayleigh wave.

Thus, in both case the same phase dependence is expected for all QDs within the laser spot. This is not observed in the experiment.